\documentclass[a4paper,12pt]{article}
\usepackage[utf8]{inputenc}
\usepackage[english]{babel}
\usepackage{amsmath}
\usepackage{amssymb}
\usepackage{graphics}
\usepackage{epsfig}
\usepackage{bbm}
\usepackage{url}
\usepackage[normalem]{ulem}
\usepackage[cyr]{aeguill}
\usepackage{geometry}
\usepackage{hyperref}
\usepackage{amsthm}
\usepackage{xcolor}

\usepackage{natbib}
\usepackage{graphicx}

\title{What does the world look like according to superdeterminism?}

\date{\textit{forthcoming in \\ the British Journal for the Philosophy of Science}}
\author{Augustin Baas and Baptiste Le Bihan \\ University of Geneva}

\begin{document}

\maketitle


\abstract{\noindent The violation of Bell inequalities seems to establish an important fact about the world: that it is non-local. However, this result relies on the assumption of the statistical independence of the measurement settings with respect to potential past events that might have determined them. Superdeterminism refers to the view that a local, and determinist, account of Bell inequalities violations is possible, by rejecting this assumption of statistical independence. We examine and clarify various problems with superdeterminism, looking in particular at its consequences on the nature of scientific laws and scientific reasoning. We argue that the view requires a neo-Humean account of at least some laws, and creates a significant problem for the use of statistical independence in other parts of physics and science more generally.}

\tableofcontents

\section{Introduction}\label{Introduction}



According to the superdeterministic view of the violation of Bell inequalities, the choices of the measures to be taken on systems are themselves determined by some past events and it is that determination that explains the violation of Bell inequalities---rather than a non-local connection existing between distant entities.  
Thus, in this approach, the violations of Bell inequalities can be explained away without positing any spooky form of action at a distance---understood in a broad sense as some sort of ontological non-locality, to be interpreted further.\footnote{In this essay, we do not say anything about how we should analyse \emph{ontologically} the concept of non-locality itself, or of non-separability, or whether several distinct relations display  non-locality in a broad sense. We shall restrict our inquiry to the possibility to adopt a local theory of quantum mechanics, in a broad sense of locality encompassing separability (see sub-section \ref{BI} for a definition).} However, this entails that the \textit{measurement independence assumption} is not satisfied. This assumption states that measurement settings are statistically independent from whatever may determine the physical state to be measured and plays a crucial role in the derivation of Bell inequalities. Superdeterministic theories have been advocated by \citet{hooft2014cellular}, \citet{palmer2016p} and \citet{HP}, and a way to test experimentally a subclass of superdeterministic theories has even been put forward by \citet{hossenfelder2011testing}. However, most physicists do not seriously consider  
superdeterministic theories as interpretations of Bell inequalities---although, importantly, they often acknowledge that the superdeterminist loophole cannot, as a matter of principle, be closed. 

In this essay, we undertake the task of clarifying and distinguishing between various reasons one may have to dismiss 
superdeterministic theories in general
(superdeterminism in the next)---with a focus on epistemological and metaphysical consequences of the view.
As we shall see, and contrary to what has unfortunately been claimed on several occasions by physicists, the strongest objections to superdeterminism can---and must---be formulated without appealing to the concept of free will. In fact, this reference to free will, which has been introduced by Bell himself, has caused some confusion during the last decades.
In the following, we first make clear why objections to
superdeterminism should avoid relying on the existence of free will, and
then we announce the structure of the paper.


Here is how Bell discussed the idea of superdeterminism, in a BBC interview in 1985 (reproduced in \citealt{davies1993ghost}):

\begin{quote}
There is a way to escape the inference of superluminal speeds and spooky action at a distance. But it involves absolute determinism in the universe, the complete absence of free will. Suppose the world is superdeterministic, with not just inanimate nature running on behind-the-scenes clockwork, but with our behavior, including our belief that we are free to choose to do one experiment rather than another, absolutely predetermined, including the `decision' by the experimenter to carry out one set of measurements rather than another, the difficulty disappears. There is no need for a faster-than-light signal to tell particle A what measurement has been carried out on particle B, because the universe, including particle A, already `knows' what that measurement, and its outcome, will be.
\end{quote}

Here we find the superdeterminist idea of an `apparent pre-agreement' between the two measurement settings and the state to be measured.
But Bell goes further and claims that \emph{free will} could not exist in a superdeterministic world. He suggest that the experimenters' capacity to freely choose the measurement settings comes under attack when operating in the background of a superdeterministic theory. Superdeterminism is hence characterised as an `absolute determinism in the universe', equated with a `complete absence of free will'. 

It is important at this stage to carefully distinguish between two assumptions about free will: (a1) free will exists and is a metaphysical prerequisite to engage with science,\footnote{Hence \cite{larsson2014loopholes}, among others, claims that ``the loophole of superdeterminism cannot be closed by scientific methods; the assumption that the world is not superdeterministic is needed to do science in the first place''. This position, in the context of Bell's tenets, originates in the works of Shimony, Horne, Clauser and Bell (for a review see \citealt{vervoort2013bell}).} and (a2) if free will exists then it guarantees, at least in some situations submitted to Bell tests, that the  measurement independence assumption is satisfied. 
This assumption states that the measurement settings are freely chosen independently of some common past event.
What Bell is not claiming here, which would constitute a third distinct assumption, is: (a3) without free will, the measurement independence assumption \emph{cannot be} satisfied. In this paper, we shall see that the first, and \textit{a fortiori} the second, assumptions are not needed in order to assess and criticise superdeterminism---and that the third assumption is simply false.

Indeed, the focus on free will strikes us as unfortunate for several reasons. First, superdeterminism is the conjunction of determinism and the atypicality of cosmological initial conditions---as we will explain in the next section---and as such, is no more problematic for free will than any deterministic theory. If one accepts the incompatibilist claim that free will is incompatible with determinism, it follows trivially that free will is incompatible with \textit{super}determinism. In fact, there are many ways to reconcile free will with determinism, namely to endorse a form of \emph{compatibilism} (for an overview, see e.g. \citealt{sep-compatibilism}). If free will is compatible with determinism, it is not clear why it should be incompatible with superdeterminism. Second, \citet{Esfeld2015bell} has recently, and convincingly to our mind, argued that Bell's theorem is logically independent from the issue of determinism--indeterminism, which entails, \textit{a fortiori}, that superdeterminism has nothing to do with free will. As we will see, the assumption of free will does not necessarily ensure the statistical independence required in Bell tests. More generally, Bell's theorem remains independent of the (super) determinism--indeterminism dispute, thereby showing that assumption (a3), namely that without free will the measurement independence assumption would not be satisfied, is false. Thus, the focus on free will is misleading: the genuine worries with superdeterminism arise from the rejection of statistical independence---not from an alleged tension with the concept of free will.

In section \ref{SD theories}, we first introduce superdeterministic theories and then assess the arguments usually given in favor of the satisfaction of the measurement independence assumption. We review the different strategies used in experimental Bell tests to select measurement settings and criticise the implicit role sometimes attributed to free will. In section \ref{Ontology SD}, we discuss the implications of superdeterminism on the metaphysics of laws of nature and in section \ref{epist} we assess the epistemological consequences that superdeterminism would have for the practice of science. 

\section{Superdeterministic Theories}\label{SD theories}

Superdeterminists reject the measurement independence assumption in sharp contrast with most other interpretations, which reject the locality assumption. In this section, we present briefly Bell inequalities, clarify how these assumptions relate to each other and close by discussing the main motivations for subscribing to the measurement independence assumption.

\subsection{Bell inequalities}\label{BI}

Bell inequalities\footnote{There are various Bell's type inequalities. In this section we consider the so-called CHSH inequality as a particular example. More generally in the paper, Bell inequalities and Bell tests refer respectively to the set of all inequalities evidencing non-locality and their tests (for an exhaustive account see \citealt{brunner2014bell}).} have been derived using different sets of assumptions and, as such, one may explain their violation in various ways depending on which assumption is rejected. Tests of Bell inequalities run as follows: measurements are performed at two different spacelike separated locations---say A and B. Call the measurement settings $x$ and $y$ respectively and the outcomes of the measurement $a$ and $b$ respectively. Call $\lambda$ some hidden variables that could have determined the measurement settings. These were historically introduced to account for the statistical nature of the quantum predictions and possibly complete the description of the quantum state, the quantum correlations being recovered by averaging over the hidden variables with some probability distribution $P$. 

The main current theories of quantum mechanics reject \emph{locality}. This is true of both deterministic approaches (including \textit{Bohmian mechanics}, see \citealt{durr2009bohmian} and the \textit{many-worlds approach}, see e.g. \citealt{wallace2012emergent}) and indeterministic approaches (with collapse theories as the so-called \textit{GRW theory}, \citealt{ghirardi1986unified}). 
But, interestingly, Bell's theorem relies more generally on the \emph{free-choice assumption}, which is already a complex assumption that must be analysed as \textit{the conjunction of the locality assumption and the measurement independence assumption}: 

\begin{itemize}
\item \textbf{Locality Assumption (LA)\footnote{It is the conjunction of the parameter independence assumption:  $P(a \vert \lambda, x,y)=P(a \vert \lambda, x)$ and $P(b \vert \lambda, x,y)=P(b \vert \lambda, y)$; and the outcome independence assumption: $P(a,b|x,y,\lambda) = P(a|x,y,\lambda) \cdot P(b|x,y,\lambda)$. See \citet[section 3.1.2]{sep-bell-theorem}.}:} $P(a,b|x,y,\lambda) = P(a|x,\lambda) \cdot P(b|y,\lambda)$.
\end{itemize}


\begin{itemize}
\item \textbf{Measurement Independence Assumption (MIA):} 
$P(x,y \vert \lambda )=P(x,y)$.
\end{itemize}

In the logical space of the possible accounts of the violation of Bell inequalities, superdeterminism exploits the possible failure of MIA. Hence, superdeterminism offers a potential local interpretation of Bell inequality violations, namely one which does not accept non-locality. However, it is important to be careful here since, as \citet{hossenfelder2011testing} rightly points out, superdeterminism also contradicts the first conjunct of the free-choice assumption, at least in some particular sense. Indeed, since the two measurement settings $x$ and $y$ are already determined via a past event, strictly speaking, there is already an indirect form of non-locality. However, this non-locality is not substantial. As Hossenfelder puts it, this non-locality ``does however a priori not necessitate superluminal exchange of information or action at a distance" \citep[1524]{hossenfelder2011testing}.  To put it in more philosophical terms, the form of non-locality entailed by superdeterminism does not rely on the existence of some \emph{primitive relation of ontological dependence} existing between the two systems, or of any other modally-loaded connecting relation (for a recent analysis of what we call `substantive non-locality', see e.g. \citeauthor{Calosi}, forthcoming). For the sake of argument, in what follows we will refer to `substantive non-locality' as `non-locality'.

Interestingly, this loophole in the inference from the violation of Bell inequalities to non-locality---since it is possible to reject MIA---cannot be definitively closed.\footnote{See also \cite{GISIN1999103}.} Assuming the causal structure of special relativity, the two events corresponding to the choices of the settings have a common past, which is given by the overlap of their backward light cones. As a result, it could be that both of them were jointly determined in the past---even \textit{before} the entangled state was produced. Therefore, and although it has been claimed that it is possible to run loophole-free Bell tests (see for instance \citealt{gallicchio2014testing,Big2018test}), there does exist a possible way to escape (substantial) non-locality, entailing a potential loophole in the inference of non-locality from the violation of Bell inequalities. 


In this context, superdeterminism refers to a class of theories that build on the rejection of MIA (see \citealt{hooft2014cellular} and \citealt{palmer2016p}).\footnote{Another possible loophole was mentioned recently by \citet{adlam2018spooky}: \emph{temporal non-locality}. Temporal non-locality shares with superdeterminism the idea that apparent cases of spatial non-locality may be explained away by considering the past. However, temporal non-locality---in shark contrast with superdeterminism---amounts to an indeterminist account of Bell inequality violations. Indeed, in this approach the causal chain of determination jumps in time over intervals, connecting timelike (rather than spacelike) separated events. It means that this view also entails the violation of statistical independence. As a result, and although we will not consider further temporal non-locality, we take our discussion of superdeterminism to be of importance for temporally non-local interpretations of quantum mechanics.} 



\subsection{The Measurement Independence Assumption}

Call $P_{mi}$ the process that generates a sequence of bits used to drive the settings. $P_{mi}$ has been realised by a pseudo-random numbers generator \citep{aspect1982experimental}, a quantum process \citep{weihs1998violation,giustina2015significant}, cosmic photons \citep{gallicchio2014testing}---a proposition implemented with photons emitted by Milky Way stars \citep{handsteiner2017cosmic}---a `cultural pseudo-random numbers' generator, that is the pixels of some digitised popular movie \citep{shalm2015strong}, or more recently the `free' instructions given by one hundred thousand people \citep{Big2018test}. These examples show different strategies, more or less explicitly acknowledged by the authors, used to give support to---allegedly---the satisfaction of MIA. 

We may identify three distinct strategies in this constellation of experiments:  

\begin{enumerate}
\item The `random outcome' strategy uses a process that we have reasons to believe to provide statistically independent outcomes: for instance, a \textit{pseudo-random numbers process}---as implemented by an algorithm running on a computer---; a \textit{quantum process} as in the physical description provided by ID quantique\footnote{See \url{https://www.idquantique.com/random-number-generation/overview/}.}; or a \textit{chaotic process}, as turbulence causing fluctuations of the transmitted intensity of a laser beam propagating through the atmosphere (see \url{random.org}). 
In this strategy, the conclusion relies  either on  statistical tests, as the ones provided by the NIST \citep{rukhin2001statistical}, or by an analysis the process generating the sequence (see 2.3). In both cases, the confidence that the outcomes are statistically independent come from the implausibility of a causal link.  

\item In the `past events' strategy, $x$ and $y$ are determined by two different events located `far away' in their past.  The strategy relies on the implausibility of a correlation between $x$ and $y$. Basically, the more the distance between the two events is, and the more the time interval between each of these events is, the more implausible a causal mechanism and thus a correlation are expected to be. \citet{handsteiner2017cosmic} used signals coming from Milky Way stars and concluded that the ``most recent time by which any local-realist influences could have engineered the observed Bell violation" cannot be earlier than around \textit{six hundred} years. This time has been claimed to be improved by around \textit{twenty} orders of magnitude by ``using pairs of quasars or patches of the cosmic microwave background'' \citep{gallicchio2014testing}.

\item Finally, the free-will strategy relies on the existence of one, or several, free agent(s) `responsible' for the choice of the measurement settings $x$ and $y$. This strategy entails a strong view about the existence of `metaphysically free' agents that can bring about some randomness within the world from the outside, so to speak. For example, \citet{gill2003comment} claim that: ``an experimenter is free to make [a choice] in the laboratory, and [...] a theoretician is free to make [a choice] in a \textit{Gedankenexperiment}'' \citep[282]{gill2003comment}. And they then go on to connect this metaphysical freedom to the existence of randomness in the world: ``We shall convert this freedom into a statistical independence assumption''.\footnote{\citet{pironio2015random} proposes different \textit{scenarios} in order to make the correlation more and more implausible; however, although not explicitly stated, it seems that each scenario falls under to the free-will strategy.}
\end{enumerate}

The first two strategies aim at assessing as highly unlikely the claim that the measurement settings have been determined in the past in such a way as to entail observed results in the present; nonetheless, and independently of the level of complexity involved in each of these strategies, none of them has the resources to definitively---as a matter of logical necessity---close the superdeterminist loophole. In a nutshell, \emph{unlikeliness is not impossibility}. The third strategy differs radically by its nature and results. It always succeeds, and that is no surprise as the strategy assumes that free will exists and (a2) (``if free will exists then it guarantees,  at least  in  some  situations  submitted  to  Bell’s  tests,  that  the  measurement independence assumption is satisfied'') without any justification. Given this \textit{petitio principii}, there is no need to test statistical independence experimentally and, importantly, it is useless to complicate the scenario by involving more and more agents or trying to combine the third strategy with one of the first two strategies---except for practical reasons as to increase the rate of instructions for the measurement settings. 
Also, it remains unclear how statistical independence could be derived from the existence of free will, and on top of that, full statistical independence is not required. Indeed, at least concerning Bell's theorem, MIA can be relaxed (see \citealt{barrett2011much} and the discussion in the next subsection). 
And regarding free will, as we pointed out in the introduction, there is no need to postulate free will when discussing the philosophical interpretations of quantum mechanics. Furthermore, the existence of free will is a highly non-trivial assumption that unnecessarily complicates the situation. Therefore, we believe that this line of thought should be discontinued. 

\subsection{Determinism and Statistical Independence} 

Statistical independence may come for free in indeterministic theories but it can also be derived within a deterministic theory. In this subsection, we first briefly sketch out the derivation of statistical independence as we find it in the literature. Then we comment on a recent work showing that, in principle, a Bell test can be conducted with the certification that the measurement independence assumption is satisfied, if cosmological initial conditions are taken to be typical. 

Statistical independence has been  derived in the case of the Galton board in  \citet{durr2009bohmian} (see also the enlightening discussions in \citealt{lazarovici2015typicality}).\footnote{A Galton board, also named a `bean machine' or `Quincunx', consists of a vertical board with horizontal lines of pegs separated by an equal distance. From one line to another, the pegs are moved from half this distance. Balls, whose diameters are smaller that the inter-pegs distances, are dropped from the top of the board and bounce either left or right as they hit the pegs, and eventually, end up in boxes at the bottom.} In what follows, we sketch out how this sort of derivation proceeds (for a formal and complete presentation, see \citealt{durr2009bohmian}). The distribution of the balls typically observed at the bottom of the Galton board fits well with a probabilistic distribution, which relies on statistical independence. To obtain these statistical regularities within a deterministic framework, we have to look at the initial randomness of the balls entering the Galton board and eventually trace back this randomness up to the beginning of the universe. In a nutshell, in order to get statistical independence in a deterministic theory---here the dynamics is assumed to be Newtonian---\citet{durr2009bohmian} assume some distribution for the initial conditions of the balls entering the Galton board, and eventually some distribution for the initial conditions of the universe, namely the \emph{cosmological initial conditions}. They then introduce a typicality measure for this set of cosmological initial conditions. Statistical independence follows from any \textit{typical} initial state.\footnote{For the sake of brevity we did not mention the `statistical hypothesis', which plays an essential role in the derivation by \citet{durr2009bohmian}: it stipulates how the typicality of the cosmological initial conditions is measured.} Think for instance about the specific situation in which the balls enter the Galton board at exactly the same location. As the dynamics is deterministic, they will all end up in the very same position, which is in contradiction with statistical regularities as usually observed. And going backward to the beginning of the universe, so to speak, this particular distribution of the balls in the Galton board can only follow from \textit{atypical} cosmological initial conditions. \textit{Therefore, the atypicality of the cosmological initial conditions does not allow to derive statistical regularities. This atypicality is precisely what superdeterminism has to 
assume in order to claim that the measurements settings are already determined by a very distant past event in spacetime.
} 

Then it is natural to ask whether a Bell test could be set up in such a way that the previous derivation could be used to satisfy MIA.\footnote{Note that there are examples of models, local and deterministic that reproduce a violation of Bell inequalities. More generally, \cite{brans1988bell} and \cite{hall2016significance} have shown that any statistical correlation has a local and deterministic model. Those models take the measurement settings to be determined by local hidden variables; they contradict MIA but they also fail to reproduce all aspects of a Bell experiments since they do not reproduce the apparent satisfaction of MIA in the experiments. In other words, those examples are not models of superdeterministic theories.} 

It is common to acknowledge that the complexity of the derivation sket-ched above makes the derivation `practically impossible' for any `realistic' system \citep{durr2009bohmian}. Now it happens that this limitation has been recently overcome, at least in principle, by using randomness amplification \citep{colbeck2012free}. In short, the idea is to relax MIA; see e.g. \cite{hall2010local,barrett2011much}. By starting with a sequence of an arbitrarily small amount of randomness, a Bell test can be used to amplify it at will, such that the sequence obtained, certified to be random, can be used to drive the measurement settings. Hence ``a relaxed free choice assumption is sufficient to establish all results derived under the assumption of virtually perfect free choices" \citep{colbeck2012free}. Now, as the randomness amplification is expressed in the quantum formalism, it can be phrased within a deterministic quantum theory, say Bohmian mechanics.  As should be clear by now, this description eventually relies on the assumption that the initial sequence includes some randomness. And, in order to take place in a deterministic framework, we need to assume the typicality of the cosmological initial conditions of such a world. What matters at this point, is that for MIA to be true in the case of the Galton board, the cosmological initial conditions must be typical. This particular case may be regarded as a model case: indeed, what is true of this particular simple case, ideal in many respects, should be expected to hold for systems more complex than the Galton board, much closer to real systems, as described in \citet{durr2009bohmian}.\footnote{This result follows strategy (a), except that the certification does not rely on statistical tests but on a derivation, which relies on the assumption of typical cosmological initial conditions.} As we shall see, the atypicality implicitly assumed by superdeterminism has strong implications for philosophical debates on the nature of laws of nature, and for the very possibility of using statistical reasoning in science.

\section{The Metaphysics of Superdeterminism}\label{Ontology SD}

Superdeterminism entails at least two interesting and potentially problematic metaphysical consequences that we review in this section. 1) It requires an \emph{atypical fine-tuning of the initial state of the universe}. 2) It entails that \emph{some laws of nature are contingent and ontologically depend on cosmological initial conditions}.


\subsection{The Fine-Tuning of Initial Conditions}

The situation bears similarities with the problem of the fine-tuning of fundamental constants in scientific cosmology. Indeed, the fundamental constants in the standard model of cosmology have highly specific values for a few parameters in such a way that, if these values had been different, the world as we know it, with its complex organisation, would not have existed, and would not have allowed for the emergence of complex systems---and, in particular, of sentient individuals. One may then ask: why is it the case that these parameters do have these \emph{specific values}? In response, physicists and philosophers have been considering various strategies such as: 1) Something unlikely to happen happened and there is nothing to be surprised about (see e.g. \citealt{juhl2006fine}); 2) A divine being created laws of nature and fine-tuned the fundamental constants in the right way to allow for life to emerge \citep{swinburne2003argument}; 3) We live in a multiverse composed of an infinity of universes with distinct values for the fundamental constants, and only some of these universes have the right values, allowing the existence of observers who then express puzzlement about the fine-tuning of the particular universe that they do inhabit (see e.g. \citealt{Smart}, \citealt{susskind}); 4) This is a false problem for some reason (for instance because the model with these parameters relies on non-fundamental theories, with the expectation that a more fundamental theory could do without such parameters, or because probabilities are ill-defined).

The situation for superdeterminism is similar but not wholly identical: the fine-tuning of initial conditions forces a choice between some of those strategies. A witty god could have fine-tuned the initial conditions, it might be that an unlikely cosmic coincidence happened, or it might be that quantum mechanics, because of its lack of fundamentality for instance, should not be used to derive philosophical consequences. However, the multiverse proposal is not available here. Atypical conditions are not required for sentient life to exist, as far as we can tell, but only for us to be able to observe systematic violations when testing Bell inequalities.

Thus, the problem of fine-tuning is not specific to superdeterminism and we will set it aside. However, as we shall see, the asymmetry between the fine-tuning of parameters in cosmology, on the one hand, and the atypicality of initial conditions in superdeterminism, on the other hand, has consequences on the metaphysics of laws of nature.

\subsection{Laws of Nature}

As said above, a slight variation in the initial conditions at the beginning of the universe would prevent systematic \emph{conclusive} Bell tests since those would require \emph{both} the violation of the inequality under scrutiny \emph{and} the experimental satisfaction of the free-choice assumption. Although the non-satisfaction of the second condition appears as a potential way out, we shall see in the next section that it conflicts with actual scientific practice. So for now, let us just assume the non-satisfaction of the first condition, i.e. that the inequalities are violated. It would entail that quantum laws theorising on those violations would not exist.\footnote{By `quantum laws' we refer to the set of laws of any theory which could give an account of the violations of Bell inequalities. The different quantum theories give examples of such sets of laws.} This means that superdeterminism entails two metaphysical consequences regarding the modal status of these laws. First, the quantum laws are \emph{contingent}---indeed, in possible worlds with different initial conditions, these laws do not exist. Second, and more problematically, these laws \emph{ontologically depend} upon the initial conditions. Let us look at the two consequences.

As a first consequence, laws of nature turn out to be contingent---or, at the very least, the \textit{quantum} laws turn out to be contingent (indeed, this contingency is logically consistent with other laws being necessary). At least, this is true if we accept that cosmological initial conditions could have been different, a claim which is not that trivial. Indeed, still in the framework of superdeterminism, if cosmological initial conditions could not have been different from what they are, then quantum laws would be necessary. Taking initial conditions as contingent is clearly a natural attitude in physics, and this is probably the mainstream position in the metaphysics of laws of nature, but it is not clear that the attitude of the physicist towards initial conditions, in practice, should be extended to \emph{cosmological} initial conditions. However, this question does not matter for our current purpose since, as we shall see, it is not the modal status of laws of nature that is problematic in the superdeterministic setting; the most pressing issue arises from the ontological dependence of laws on cosmological initial conditions.


Superdeterminism entails that the cosmological initial conditions \emph{shape} the form of the laws of nature or, as mentioned above, of at least the quantum laws.\footnote{Cf. the previous note.} This view is quite unusual and should be noted since laws of nature are commonly regarded as being ontologically independent from any set of initial conditions, in the sense that varying the initial conditions do not modify the laws of nature.\footnote{With the interesting exception of the neo-Humean view, as we shall see below.} 
Superdeterminism, however entails that quantum laws ontologically depend on initial conditions in general and, \textit{in fine}, on the cosmological initial conditions. This consequence is quite \emph{strange} and sheds suspicion on the whole idea of superdeterminism. But, as we shall see: a) this picture looks more reasonable if we adopt a neo-Humean picture of laws and, b) more generally, our intuition that laws of nature cannot depend upon the cosmological initial conditions follows from a particular account of laws of nature---namely, \emph{primitivism about laws}.




There are mainly four views about the nature of law: 1) \emph{primitivism}, namely the view that laws of nature are primitive entities that cannot be explained further (see e.g.  \citealt{tim2007metaphysics}), 2) the \emph{neo-Humean view} championed by \citet[ix]{lewis1986plurality} that ``all there is in the world is a vast mosaic of local matters of particular fact, just one little thing and then another''; 3) the \emph{dispositionalist view} that identify laws of nature to dispositional properties of objects, processes, natural kinds and/or other categories of entities \citep{ellis2001scientific, bird2005dispositionalist} and 4) the \emph{DTA view} that identify laws of nature to second-order relations of necessitation connecting first-order universals, a view defended by \citet{dretske1977laws}, \citet{tooley1977nature} and \citet{armstrong1978theory, armstrong2016law}.


The clean distinction between laws and initial conditions somewhat vanishes in the context of a \emph{neo-Humean picture}, since laws of nature do not exist in this account. What we do have at the ontological level is only a mosaic of facts, properties, or events, organised by a background ordering structure, standardly identified with spacetime. There is some room for interpreting the exact nature of the \emph{organised building blocks} (facts, properties, objects, events) and the exact nature of the \emph{ordering structure} (a substantial spacetime, a relationist spacetime, a metric field coupled to a derivative manifold of points, a metric field only, or a non-spatiotemporal quantum gravity structure like spin foams for instance\footnote{The nature of spacetime varies from one interpretation of GR to another, and is interpreted differently in the various approaches to quantum gravity (see e.g. \citealt{HW} and \citealt{BLBNL}).}); but the general point that matters here is the non-modal character of the building blocks and the ordering structure. The distinction between initial conditions and laws is then moved from the ontology to the linguistic descriptions: general statements from which we may derive a lot of particular statements, because they target regularities at the ontological level, are scientific laws. In this picture, the fact that some regularities `depend upon' the initial conditions is not particularly problematic. Bell inequality violations follow from a strange coincidence, that we regard as a law of nature; but this is just the general strategy of the neo-Humean regarding laws of nature, since there is no external device of necessitation constraining the distribution of entities in the natural world.

\emph{Primitivism}, \emph{the dispositionalist view} and \emph{the DTA view}, three particular versions of a broader necessitarian view, on the other hand, come into conflict with superdeterminism. Indeed, these accounts do rely on a sharp distinction between \emph{dynamical laws} and the \emph{systems they evolve}. The ontological dependency of laws on cosmological initial conditions becomes puzzling since these laws are something over and above the entities they are evolving. How is a necessitarian superdeterminist to explain that the necessitation device (primitive laws, dispositional properties or necessitation relations) depends for its existence on the highly specific cosmological initial conditions of the actual world? Avoiding cosmic coincidence is usually regarded as one of the main reasons to adopt a form of necessitarianism against the neo-Humean picture. A necessitarian superdeterminist would loose all the benefit of necessitarianism.\footnote{Note that a way out for a primitivist superdeterminist would be to state that the so-called quantum laws are not laws after all; but then the burden of the proof would be on the shoulders of the advocate of such a move: what would be the relevant criterion to trust other nomological statements to actually refer to genuine laws of nature?} 
As a consequence, superdeterminism strongly motivates endorsing the neo-Humean view. This point might count as a bad or a good thing depending on the reader's allegiance to the necessitarian or the neo-Humean side. Therefore, superdeterminism undermines the main motivations for most account of laws of nature. However, the neo-Humean view offers an interesting way out for the superdeterminist. As we will now see, the more serious issues with superdeterminism are epistemological.



\section{The Epistemology of Superdeterminism}\label{epist}

In this section, we review two epistemological arguments against superdeterminism. Before that, let us make a comment on a potential weakness of superdeterminism sometimes noticed. Superdeterminism is not a theory yet, but rather an ensemble of propositions based on the possibility to formulate a local and deterministic theory, compatible with the violation of Bell inequalities. To be a theory, superdeterminism would have to provide an explanation of the exact values associated with the violations of Bell inequalities, which is what quantum mechanics does.\footnote{That is another  expected task for a theory of superdetermimism. 
If it is possible in principle to formulate a local and deterministic theory that accounts for the violation of Bell inequalities, it is unclear yet how it could impose any bound, except the algebraic bound of 4, contrary to quantum mechanics, which in the case of CHSH inequality is limited by Tsirelson's bound $2\sqrt{2}$ (see \citealt{sep-bell-theorem}).}

We therefore propose more direct epistemological arguments against superdeterminism in this section.
First, one might claim that superdeterministic theories are empirically incoherent by being at odds with their empirical evidence. Second, statistical independence plays an essential role in contemporary science, and cannot be dismissed without loosing the epistemological justification of science as a whole. 


\subsection{The Problem of Empirical Coherence}


Issues of empirical coherence have been introduced by \cite{barrett} in the context of quantum mechanics. As he writes: ``In order to judge whether a theory is empirically adequate one must have epistemic access to reliable records of past measurement results that can be compared against the predictions of the theory'' \cite[49]{barrett}. He then distinguishes theories of quantum mechanics that can pass the test from other theories of quantum mechanics that cannot pass this test. Problems of empirical coherence have then been found to be pervasive in physics with the \emph{configuration space realist interpretation of quantum mechanics},\footnote{According to this approach to quantum mechanics, the wave function is a real entity, and since it is defined on a configuration space rather than the ordinary 3D space, we should accept that our physical world is a physical counterpart of the configuration space---not a 3D space. A problem of empirical coherence is then to understand the connection between the fundamental physical structure and the 3D world, and to explain how evidence apparently taking place in 3D world can justify the view that we live in a structure made of a huge number of dimensions. See e.g. \cite{monton2002}, \cite{maudlin2007completeness},  \cite{albert2013} and \cite{BLBdisputatio}.} and \emph{quantum gravity}.\footnote{\cite{HW} have argued that the potential disappearance of space and time, as observed in various approaches to quantum gravity, leads to an interesting issue of empirical coherence: how is it possible to justify a theory claiming that space and time do not exist with evidence localised in space and time?}

Superdeterminists have to deal with their own novel brand of empirical coherence, one which is at least as problematic as the sorts of empirical coherence to be found in configuration space realism and quantum gravity. Indeed, on the one hand, superdeterminists claim that we should let go MIA; on the other hand they assume MIA in order to interpret experiments. Indeed, as we will discuss in the next section in more detail, statistical data are produced and regarded as being reliable because they were produced operating under the assumption of MIA. To put it differently, superdeterminists reject MIA 
in order to make sense of a result obtained by running experiments and interpreting outputs of those experiments by appealing to statistical independence. This incoherence is empirical in that it is central to the way empirical knowledge, based on statistical analysis, is produced. With the empirical incoherence associated with the denial of the existence of spacetime, the issue was with each observation/experiment taken separately; with superdeterminism the empirical incoherence appears at a different level when using collections of observations to draw consequences. This means that superdeterminists---by rejecting an assumption essential to the creation of data to be analysed later on---commit a \emph{dialectical mistake}.

\subsection{Reasoning in Science}

The second epistemological argument against superdeterministic theories starts with the realisation that on a superdeterministic view scientists can \textit{never} assume statistical independence. We follow here \cite{goldstein2011bell} when they write:

\begin{quote}
[T]his assumption is necessarily always made whenever one does any empirical science; in practice, one assesses the applicability of the assumption to a given experiment by examining the care with which the experimental design precludes any non-conspiratorial dependencies between the preparation of the systems and the settings of instruments. 
[...]

[I]f you are performing a drug versus placebo clinical trial, then you have to select some group of patients to get the drug and some group of patients to get the placebo. The conclusions drawn from the study will necessarily depend on the assumption that the method of selection is independent of whatever characteristics those patients might have that might influence how they react to the drug. \citep{goldstein2011bell}
\end{quote}





Furthermore, the denial of statistical independence conflicts with basic notions that allows for scientific practice---both theoretical and experimental. Indeed, statistical independence is a necessary assumption when doing science---one which is essential if one wants to make sense of notions such as \emph{isolated system}, \emph{repetition of an experiment} or \emph{random measurement error}. 

Rejecting the statistical independence of the successive runs of experiments restricts drastically the possibility to describe systems as isolated systems---strictly speaking, only the whole universe fulfills the condition of isolation. It might be so, but again, to assume that the dynamics is deterministic does not prevent statistical independence to exist and does not bar the road to describing sub-systems as being isolated. Furthermore, rejecting statistical independence goes against common practice in modern science. Indeed, when repeating an experiment, it is usually assumed that the successive experimental runs are independent from each other. Thus, if we do not assume statistical independence to begin with, comparing those experiments turns into an impossible task. 


Therefore, if we reject statistical independence as a whole, it becomes impossible to run repeated measurements since we may no longer suppose that runs are independent from each other. If the runs are not independent, it is not possible anymore to compare them and, so, to offer statistical interpretations of the data. In a slogan, a full-blown rejection of statistical independence dooms statistical science. At this point one may wonder: should a superdeterminist really subscribe to a full-blown rejection of statistical independence? Why not just adopt the more moderate view that statistical independence only admits of some exceptions?


According to this approach, MIA fails to apply in---\textit{and only in}---some specific circumstances: when Bell inequalities are violated. Let us call this interpretation `exceptionalist superdeterminism' (exceptionalist SD hereafter). One serious challenge for exceptionalist SD is then to understand how statistical independence might be sometimes satisfied, and sometimes not, without being at odds with the rejection of MIA in some contexts. Indeed, an exceptionalist SD must acknowledge that statistical independence applies \emph{in some circumstances}. Thus, they must assert the impossibility of using those systems in order to set the measurement settings---namely, to implement $P_{mi}$. But what ground is there for such an impossibility? One first option is that there exists some kind of cosmic principle, a fundamental law or a hand of God, preventing us of using those systems to set the measurement settings. A second option is that systems usually obeying to statistical independence cease to obey statistical independence as soon as they are put to work to set the measurement settings. These two options are both \textit{ad hoc} and unattractive. We conclude that exceptionalist SD is a dead end and that superdeterminists cannot appeal to violations \textit{à la carte} of statistical independence.

To conclude this section, let's take a step back and comment on the distinction between determinism and superdeterminism. It is common to find claims in the literature about the alleged incapacity of deterministic theories to handle statistical independence. But this is not true, as many important works in the Bohmian tradition have shown, and determinism in itself has nothing to do with the rejection of statistical independence. Superdeterministic theories---a subset of deterministic theories---take the further step of rejecting statistical independence, with the unpalatable consequences that were discussed in this paper. We hope that this work helps to bring into the light the specific issues of superdeterminism, to which deterministic theories such as Newtonian mechanics or Bohmian mechanics must not be associated with. 

\section{Conclusion}

Superdeterminism is a strange interpretation of the violation of Bell inequalities. It states that the world we live in is an incredible coincidence, entailing the existence of so-called quantum laws, which turn out to be contingent and ontologically depend upon the cosmological initial conditions. The account constrains how we should construe laws of nature as it is incompatible with the view that all laws of nature are external `entities' evolving material systems through time. This is one bullet the superdeterminist might be willing to bite. Another issue is that their view entails a global failure of the principle of statistical independence, a principle used virtually and successfully everywhere in contemporary science. Although not necessarily a damning issue for superdeterminism, it is far from clear how the violations of statistical independence in quantum mechanics can be stopped from propagating to classical physical systems: the proponent of superdeterminism should provide such an explanation in order to connect superdeterminism to the rest of science.
We also note that the original motivation for superdeterminism---saving locality---is not fully present in the picture of the world we get from it. Indeed, superdeterminism entails a form of holism as everything remains `connected' in a weak sense since the building blocks of the world cannot be isolated and probed independently of each other.

\section*{Acknowledgments} For helpful comments on an earlier draft of this essay, we would like to thank Alastair Abbott, Cyril Branciard, Nicolas Gisin, Flavien Hirsch, Rasmus Jaksland, Niels Linnemann, Christian Wüthrich and two anonymous reviewers. This work was supported by the Swiss National Science Foundation.





 

\bibliographystyle{chicago}
\bibliography{references}

\begin{thebibliography}{}

\bibitem[\protect\citeauthoryear{Abell\'an}{Abell\'an}{2018}]{Big2018test}
Abell\'an, C. (2018).
\newblock Challenging local realism with human choices.
\newblock {\em Nature\/}~{\em 557}, 212.

\bibitem[\protect\citeauthoryear{Adlam}{Adlam}{2018}]{adlam2018spooky}
Adlam, E. (2018).
\newblock Spooky action at a temporal distance.
\newblock {\em Entropy\/}~{\em 20\/}(1), 41.

\bibitem[\protect\citeauthoryear{Albert}{Albert}{2013}]{albert2013}
Albert, D.~Z. (2013).
\newblock Wave function realism.
\newblock In A.~Ney and D.~Z. Albert (Eds.), {\em The Wave Function: Essays on
  the Metaphysics of Quantum Mechanics}, pp.\  52--57. Oxford University Press.

\bibitem[\protect\citeauthoryear{Armstrong}{Armstrong}{1978}]{armstrong1978theory}
Armstrong, D.~M. (1978).
\newblock {\em A Theory of Universals}, Volume~2.
\newblock Cambridge: Cambridge University Press.

\bibitem[\protect\citeauthoryear{Armstrong}{Armstrong}{1983}]{armstrong2016law}
Armstrong, D.~M. (1983).
\newblock {\em What is a Law of Nature?}
\newblock Cambridge: Cambridge University Press.

\bibitem[\protect\citeauthoryear{Aspect, Dalibard, and Roger}{Aspect
  et~al.}{1982}]{aspect1982experimental}
Aspect, A., J.~Dalibard, and G.~Roger (1982).
\newblock Experimental test of {B}ell's inequalities using time-varying
  analyzers.
\newblock {\em Physical Review Letters\/}~{\em 49\/}(25), 1804.

\bibitem[\protect\citeauthoryear{Barrett and Gisin}{Barrett and
  Gisin}{2011}]{barrett2011much}
Barrett, J. and N.~Gisin (2011).
\newblock How much measurement independence is needed to demonstrate
  nonlocality?
\newblock {\em Physical Review Letters\/}~{\em 106\/}(10), 100406.

\bibitem[\protect\citeauthoryear{Barrett}{Barrett}{1996}]{barrett}
Barrett, J.~A. (1996).
\newblock Empirical adequacy and the availability of reliable records in
  quantum mechanics.
\newblock {\em Philosophy of Science\/}~{\em 63\/}(1), 49--64.

\bibitem[\protect\citeauthoryear{Bird}{Bird}{2005}]{bird2005dispositionalist}
Bird, A. (2005).
\newblock The dispositionalist conception of laws.
\newblock {\em Foundations of Science\/}~{\em 10\/}(4), 353--370.

\bibitem[\protect\citeauthoryear{Brans}{Brans}{1988}]{brans1988bell}
Brans, C.~H. (1988).
\newblock Bell's theorem does not eliminate fully causal hidden variables.
\newblock {\em International Journal of Theoretical Physics\/}~{\em 27\/}(2),
  219--226.

\bibitem[\protect\citeauthoryear{Brunner, Cavalcanti, Pironio, Scarani, and
  Wehner}{Brunner et~al.}{2014}]{brunner2014bell}
Brunner, N., D.~Cavalcanti, S.~Pironio, V.~Scarani, and S.~Wehner (2014).
\newblock Bell nonlocality.
\newblock {\em Reviews of Modern Physics\/}~{\em 86\/}(2), 419.

\bibitem[\protect\citeauthoryear{Calosi and Morganti}{Calosi and
  Morganti}{2018}]{Calosi}
Calosi, C. and M.~Morganti (2018, 09).
\newblock {Interpreting Quantum Entanglement: Steps towards Coherentist Quantum
  Mechanics}.
\newblock {\em The British Journal for the Philosophy of Science\/}.
\newblock axy064.

\bibitem[\protect\citeauthoryear{Colbeck and Renner}{Colbeck and
  Renner}{2012}]{colbeck2012free}
Colbeck, R. and R.~Renner (2012).
\newblock Free randomness can be amplified.
\newblock {\em Nature Physics\/}~{\em 8\/}(6), 450.

\bibitem[\protect\citeauthoryear{Davies and Brown}{Davies and
  Brown}{1993}]{davies1993ghost}
Davies, P. C.~W. and J.~R. Brown (1993).
\newblock {\em The Ghost in the Atom: A Discussion of the Mysteries of Quantum
  Physics}.
\newblock Cambridge: Cambridge University Press.

\bibitem[\protect\citeauthoryear{Dretske}{Dretske}{1977}]{dretske1977laws}
Dretske, F. (1977).
\newblock Laws of nature.
\newblock {\em Philosophy of Science\/}~{\em 44\/}(2), 248--268.

\bibitem[\protect\citeauthoryear{D{\"u}rr and Teufel}{D{\"u}rr and
  Teufel}{2009}]{durr2009bohmian}
D{\"u}rr, D. and S.~Teufel (2009).
\newblock {\em Bohmian Mechanics: the Physics and Mathematics of Quantum
  Theory}.
\newblock Springer.

\bibitem[\protect\citeauthoryear{Ellis}{Ellis}{2001}]{ellis2001scientific}
Ellis, B. (2001).
\newblock {\em Scientific Essentialism}.
\newblock Cambridge: Cambridge University Press.

\bibitem[\protect\citeauthoryear{Esfeld}{Esfeld}{2015}]{Esfeld2015bell}
Esfeld, M. (2015).
\newblock Bell's theorem and the issue of determinism and indeterminism.
\newblock {\em Foundations of Physics\/}~{\em 45\/}(5), 471--482.

\bibitem[\protect\citeauthoryear{Gallicchio, Friedman, and Kaiser}{Gallicchio
  et~al.}{2014}]{gallicchio2014testing}
Gallicchio, J., A.~S. Friedman, and D.~I. Kaiser (2014).
\newblock Testing {B}ell’s inequality with cosmic photons: Closing the
  setting-independence loophole.
\newblock {\em Physical Review Letters\/}~{\em 112\/}(11), 110405.

\bibitem[\protect\citeauthoryear{Ghirardi, Rimini, and Weber}{Ghirardi
  et~al.}{1986}]{ghirardi1986unified}
Ghirardi, G.~C., A.~Rimini, and T.~Weber (1986).
\newblock Unified dynamics for microscopic and macroscopic systems.
\newblock {\em Physical Review D\/}~{\em 34\/}(2), 470.

\bibitem[\protect\citeauthoryear{Gill, Weihs, Zeilinger, and {\.Z}ukowski}{Gill
  et~al.}{2003}]{gill2003comment}
Gill, R., G.~Weihs, A.~Zeilinger, and M.~{\.Z}ukowski (2003).
\newblock Comment on `{E}xclusion of time in the theorem of {B}ell' by {K. Hess
  and W. Philipp}.
\newblock {\em EPL (Europhysics Letters)\/}~{\em 61\/}(2), 282.

\bibitem[\protect\citeauthoryear{Gisin and Zbinden}{Gisin and
  Zbinden}{1999}]{GISIN1999103}
Gisin, N. and H.~Zbinden (1999).
\newblock Bell inequality and the locality loophole: Active versus passive
  switches.
\newblock {\em Physics Letters A\/}~{\em 264\/}(2), 103--107.

\bibitem[\protect\citeauthoryear{Giustina, Versteegh, Wengerowsky, Handsteiner,
  Hochrainer, Phelan, Steinlechner, Kofler, Larsson, Abell{\'a}n,
  et~al.}{Giustina et~al.}{2015}]{giustina2015significant}
Giustina, M., M.~A. Versteegh, S.~Wengerowsky, J.~Handsteiner, A.~Hochrainer,
  K.~Phelan, F.~Steinlechner, J.~Kofler, J.-{\AA}. Larsson, C.~Abell{\'a}n,
  et~al. (2015).
\newblock Significant-loophole-free test of {B}ell’s theorem with entangled
  photons.
\newblock {\em Physical Review Letters\/}~{\em 115\/}(25), 250401.

\bibitem[\protect\citeauthoryear{Goldstein, Norsen, Tausk, and
  Zangh{\`\i}}{Goldstein et~al.}{2011}]{goldstein2011bell}
Goldstein, S., T.~Norsen, D.~V. Tausk, and N.~Zangh{\`\i} (2011).
\newblock Bell's theorem.
\newblock {\em Scholarpedia\/}~{\em 6\/}(10), 8378.

\bibitem[\protect\citeauthoryear{Hall}{Hall}{2010}]{hall2010local}
Hall, M.~J. (2010).
\newblock Local deterministic model of singlet state correlations based on
  relaxing measurement independence.
\newblock {\em Physical Review Letters\/}~{\em 105\/}(25), 250404.

\bibitem[\protect\citeauthoryear{Hall}{Hall}{2016}]{hall2016significance}
Hall, M.~J. (2016).
\newblock The significance of measurement independence for bell inequalities
  and locality.
\newblock In T.~Asselmeyer-Maluga (Ed.), {\em At the Frontier of Spacetime :
  Scalar-tensor Theory, Bells Inequality, Machs Principle, Exotic Smoothness},
  Volume 183, pp.\  189--204. Springer.

\bibitem[\protect\citeauthoryear{Handsteiner, Friedman, Rauch, Gallicchio, Liu,
  Hosp, Kofler, Bricher, Fink, Leung, et~al.}{Handsteiner
  et~al.}{2017}]{handsteiner2017cosmic}
Handsteiner, J., A.~S. Friedman, D.~Rauch, J.~Gallicchio, B.~Liu, H.~Hosp,
  J.~Kofler, D.~Bricher, M.~Fink, C.~Leung, et~al. (2017).
\newblock Cosmic {B}ell test: {M}easurement settings from {M}ilky {W}ay stars.
\newblock {\em Physical Review Letters\/}~{\em 118\/}(6), 060401.

\bibitem[\protect\citeauthoryear{Hossenfelder}{Hossenfelder}{2011}]{hossenfelder2011testing}
Hossenfelder, S. (2011).
\newblock Testing super-deterministic hidden variables theories.
\newblock {\em Foundations of Physics\/}~{\em 41\/}(9), 1521.

\bibitem[\protect\citeauthoryear{Hossenfelder and Palmer}{Hossenfelder and
  Palmer}{2020}]{HP}
Hossenfelder, S. and T.~Palmer (2020).
\newblock Rethinking superdeterminism.
\newblock {\em Frontiers in Physics\/}~{\em 8}, 139.

\bibitem[\protect\citeauthoryear{Huggett and W{\"u}thrich}{Huggett and
  W{\"u}thrich}{2013}]{HW}
Huggett, N. and C.~W{\"u}thrich (2013).
\newblock Emergent spacetime and empirical (in) coherence.
\newblock {\em Studies in History and Philosophy of Modern Physics\/}~{\em
  44\/}(3), 276--285.

\bibitem[\protect\citeauthoryear{Juhl}{Juhl}{2006}]{juhl2006fine}
Juhl, C. (2006).
\newblock Fine-tuning is not surprising.
\newblock {\em Analysis\/}~{\em 66\/}(4), 269--275.

\bibitem[\protect\citeauthoryear{Larsson}{Larsson}{2014}]{larsson2014loopholes}
Larsson, J.-{\AA}. (2014).
\newblock Loopholes in bell inequality tests of local realism.
\newblock {\em Journal of Physics A: Mathematical and Theoretical\/}~{\em
  47\/}(42), 424003.

\bibitem[\protect\citeauthoryear{Lazarovici and Reichert}{Lazarovici and
  Reichert}{2015}]{lazarovici2015typicality}
Lazarovici, D. and P.~Reichert (2015).
\newblock Typicality, irreversibility and the status of macroscopic laws.
\newblock {\em Erkenntnis\/}~{\em 80\/}(4), 689--716.

\bibitem[\protect\citeauthoryear{Le~Bihan}{Le~Bihan}{2018}]{BLBdisputatio}
Le~Bihan, B. (2018).
\newblock Space emergence in contemporary physics: Why we do not need
  fundamentality, layers of reality and emergence.
\newblock {\em Disputatio\/}~{\em 10\/}(49), 71--95.

\bibitem[\protect\citeauthoryear{Le~Bihan and Linnemann}{Le~Bihan and
  Linnemann}{2019}]{BLBNL}
Le~Bihan, B. and N.~Linnemann (2019).
\newblock Have we lost spacetime on the way? {N}arrowing the gap between
  general relativity and quantum gravity.
\newblock {\em Studies in History and Philosophy of Modern Physics\/}~{\em 65},
  112--121.

\bibitem[\protect\citeauthoryear{Lewis}{Lewis}{1986}]{lewis1986plurality}
Lewis, D. (1986).
\newblock {\em On the Plurality of Worlds}.
\newblock Blackwell.

\bibitem[\protect\citeauthoryear{Maudlin}{Maudlin}{2007a}]{tim2007metaphysics}
Maudlin, T. (2007a).
\newblock {\em The Metaphysics Within Physics}.
\newblock Oxford University Press.

\bibitem[\protect\citeauthoryear{Maudlin}{Maudlin}{2007b}]{maudlin2007completeness}
Maudlin, T.~W. (2007b).
\newblock Completeness, supervenience and ontology.
\newblock {\em Journal of Physics A: Mathematical and Theoretical\/}~{\em
  40\/}(12), 3151.

\bibitem[\protect\citeauthoryear{McKenna and Coates}{McKenna and
  Coates}{2016}]{sep-compatibilism}
McKenna, M. and D.~J. Coates (2016).
\newblock Compatibilism.
\newblock In E.~N. Zalta (Ed.), {\em The Stanford Encyclopedia of Philosophy\/}
  (Winter 2016 ed.). Metaphysics Research Lab, Stanford University.

\bibitem[\protect\citeauthoryear{Monton}{Monton}{2002}]{monton2002}
Monton, B. (2002).
\newblock Wave function ontology.
\newblock {\em Synthese\/}~{\em 130\/}(2), 265--277.

\bibitem[\protect\citeauthoryear{Myrvold, Genovese, and Shimony}{Myrvold
  et~al.}{2019}]{sep-bell-theorem}
Myrvold, W., M.~Genovese, and A.~Shimony (2019).
\newblock {Bell’s Theorem}.
\newblock In E.~N. Zalta (Ed.), {\em The {Stanford} Encyclopedia of
  Philosophy\/} (Spring 2019 ed.). Metaphysics Research Lab, Stanford
  University.

\bibitem[\protect\citeauthoryear{Palmer}{Palmer}{2016}]{palmer2016p}
Palmer, T. (2016).
\newblock p-adic distance, finite precision and emergent superdeterminism: A
  number-theoretic consistent-histories approach to local quantum realism.
\newblock {\em arXiv preprint arXiv:1609.08148\/}.

\bibitem[\protect\citeauthoryear{Pironio}{Pironio}{2015}]{pironio2015random}
Pironio, S. (2015).
\newblock Random `choices' and the locality loophole.
\newblock {\em arXiv preprint arXiv:1510.00248\/}.

\bibitem[\protect\citeauthoryear{Rukhin, Soto, Nechvatal, Smid, and
  Barker}{Rukhin et~al.}{2001}]{rukhin2001statistical}
Rukhin, A., J.~Soto, J.~Nechvatal, M.~Smid, and E.~Barker (2001).
\newblock A statistical test suite for random and pseudorandom number
  generators for cryptographic applications.
\newblock Technical report, Booz-Allen and Hamilton Inc Mclean Va.

\bibitem[\protect\citeauthoryear{Shalm, Meyer-Scott, Christensen, Bierhorst,
  Wayne, Stevens, Gerrits, Glancy, Hamel, Allman, et~al.}{Shalm
  et~al.}{2015}]{shalm2015strong}
Shalm, L.~K., E.~Meyer-Scott, B.~G. Christensen, P.~Bierhorst, M.~A. Wayne,
  M.~J. Stevens, T.~Gerrits, S.~Glancy, D.~R. Hamel, M.~S. Allman, et~al.
  (2015).
\newblock Strong loophole-free test of local realism.
\newblock {\em Physical Review Letters\/}~{\em 115\/}(25), 250402.

\bibitem[\protect\citeauthoryear{Smart}{Smart}{1989}]{Smart}
Smart, J. (1989).
\newblock {\em Our Place in the Universe: A Metaphysical Discussion}.
\newblock Oxford: Blackwell.

\bibitem[\protect\citeauthoryear{Susskind}{Susskind}{2005}]{susskind}
Susskind, L. (2005).
\newblock {\em The Cosmic Landscape: String Theory and the Illusion of
  Intelligent Design}.
\newblock Back Bay Books.

\bibitem[\protect\citeauthoryear{Swinburne}{Swinburne}{2003}]{swinburne2003argument}
Swinburne, R. (2003).
\newblock The argument to {G}od from fine-tuning reassessed.
\newblock In N.~A. Manson (Ed.), {\em God and Design}, pp.\  121--139.
  Routledge.

\bibitem[\protect\citeauthoryear{'t~Hooft}{'t~Hooft}{2014}]{hooft2014cellular}
't~Hooft, G. (2014).
\newblock The cellular automaton interpretation of quantum mechanics.
\newblock {\em arXiv preprint arXiv:1405.1548\/}.

\bibitem[\protect\citeauthoryear{Tooley}{Tooley}{1977}]{tooley1977nature}
Tooley, M. (1977).
\newblock The nature of laws.
\newblock {\em Canadian Journal of Philosophy\/}~{\em 7\/}(4), 667--698.

\bibitem[\protect\citeauthoryear{Vervoort}{Vervoort}{2013}]{vervoort2013bell}
Vervoort, L. (2013).
\newblock Bell's theorem: Two neglected solutions.
\newblock {\em Foundations of Physics\/}~{\em 43\/}(6), 769--791.

\bibitem[\protect\citeauthoryear{Wallace}{Wallace}{2012}]{wallace2012emergent}
Wallace, D. (2012).
\newblock {\em The Emergent Multiverse: Quantum Theory According to the Everett
  Interpretation}.
\newblock Oxford: Oxford University Press.

\bibitem[\protect\citeauthoryear{Weihs, Jennewein, Simon, Weinfurter, and
  Zeilinger}{Weihs et~al.}{1998}]{weihs1998violation}
Weihs, G., T.~Jennewein, C.~Simon, H.~Weinfurter, and A.~Zeilinger (1998).
\newblock Violation of {B}ell's inequality under strict {E}instein locality
  conditions.
\newblock {\em Physical Review Letters\/}~{\em 81\/}(23), 5039.

\end{thebibliography}
\end{document}